\documentclass[twocolumn,showpacs,preprintnumbers,amsmath,amssymb]{revtex4}

\usepackage{graphicx}
\usepackage{dcolumn}
\usepackage{bm}
\usepackage[colorlinks]{hyperref}
\newcounter{foo}
\def\RomanSuji#1{\setcounter{foo}{#1}\Roman{foo}}

\begin{document}


\title{Recoverable prevalence in growing scale-free networks
\\ and the effective immunization}

\author{Yukio Hayashi}
\author{Masato Minoura}
\author{Jun Matsukubo}
\affiliation{%
Japan Advanced Institute of Science and Technology,\\
Ishikawa, 923-1292, Japan
}%


\date{\today}

\begin{abstract}
We study the persistent recoverable prevalence and the extinction of
 computer viruses 
via e-mails on a growing scale-free network with new users,
which structure is estimated form real data.
The typical phenomenon is simulated in a realistic model with the 
 probabilistic execution and detection of viruses.
Moreover, the conditions of extinction by random and targeted 
 immunizations for hubs are derived through bifurcation analysis for
 simpler models by using a mean-field approximation without the
 connectivity correlations.
We can qualitatively understand the mechanisms of the spread in
 linearly growing scale-free networks.
\end{abstract}

\maketitle


\section{INTRODUCTION}
In spite of the different social, technological, and biological
interactions, many complex networks in real-worlds have a common
structure based on the universal self-organized mechanism:
network growth and preferential attachment of connections
\cite{Albert01}\cite{Barabasi99}.
The structure is called scale-free (SF) network, which 
exhibits a power-law degree distribution 
$P(k) \sim k^{- \gamma}$, $2 < \gamma < 3$,
for the probability of $k$ connections.
The topology deviates
from the conventional homogeneous regular lattices and random graphs.
Many researchers are attracted to a new paradigm of the
heterogeneous SF networks in the active and fruitful area.

The structure of SF networks 
also gives us a strong impact on the dynamics of 
epidemic models for computer viruses, HIV, and others.
Recently, it has been shown \cite{Satorras01a} that
a susceptible-infected-susceptible (SIS) model on SF networks has
no epidemic threshold; infections can be proliferated, whatever
small infection rate they have.
This result disproves the threshold theory in
epidemiology \cite{Shigesada97}.
The heterogeneous structure is also crucial for spreading the
viruses on the analysis of susceptible-infected-recovered (SIR)
models \cite{May01}\cite{Newman02a}.
In contrast to the absence of epidemic threshold,
an immunization strategy has been theoretically presented 
in SIS models \cite{Dezso02}\cite{Satorras02}.
The targeted immunization applies 
the extreme disconnections by attacks against hubs with high-degrees 
on SF networks \cite{Albert00a}
to a prevention against the spread of infections.

In this paper, we 
investigate the dynamic properties for spreading of computer viruses 
on the SF networks estimated from real data of e-mail
communication \cite{Mikami01}.
As a new property in both simulation and theoretical analysis, 
we suggest a growing network with new e-mail users causes 
the recoverable prevalence from a temporary silence of 
almost complete extinction.
The typical phenomenon in observations
\cite{Kephart93b}\cite{White95} is 
not explained by the above statistical 
analysises at steady states or mean-values 
(in the fixed size or $N \rightarrow \infty$).
We first consider, in simulations, 
a realistic growing 
model with the probabilistic execution and detection of
viruses on the SF network.
Then, for understanding the mechanisms of recoverable
prevalence and extinction, we analyze simpler 
growing models in deterministic equations.
By using a mean-field approximation without the connectivity
correlations, 
we derive bifurcation conditions from the extinction to the 
recoverable prevalence (or the opposite),
which is related to the growth, infection, and immune rates.
Moreover, we verify the effectiveness of the targeted immunization
by anti-viruses for hubs 
even in the growing system.

\section{E-MAIL NETWORK}
\subsection{The state transition for infection}
\begin{figure}[htb]
\includegraphics[width=70mm]{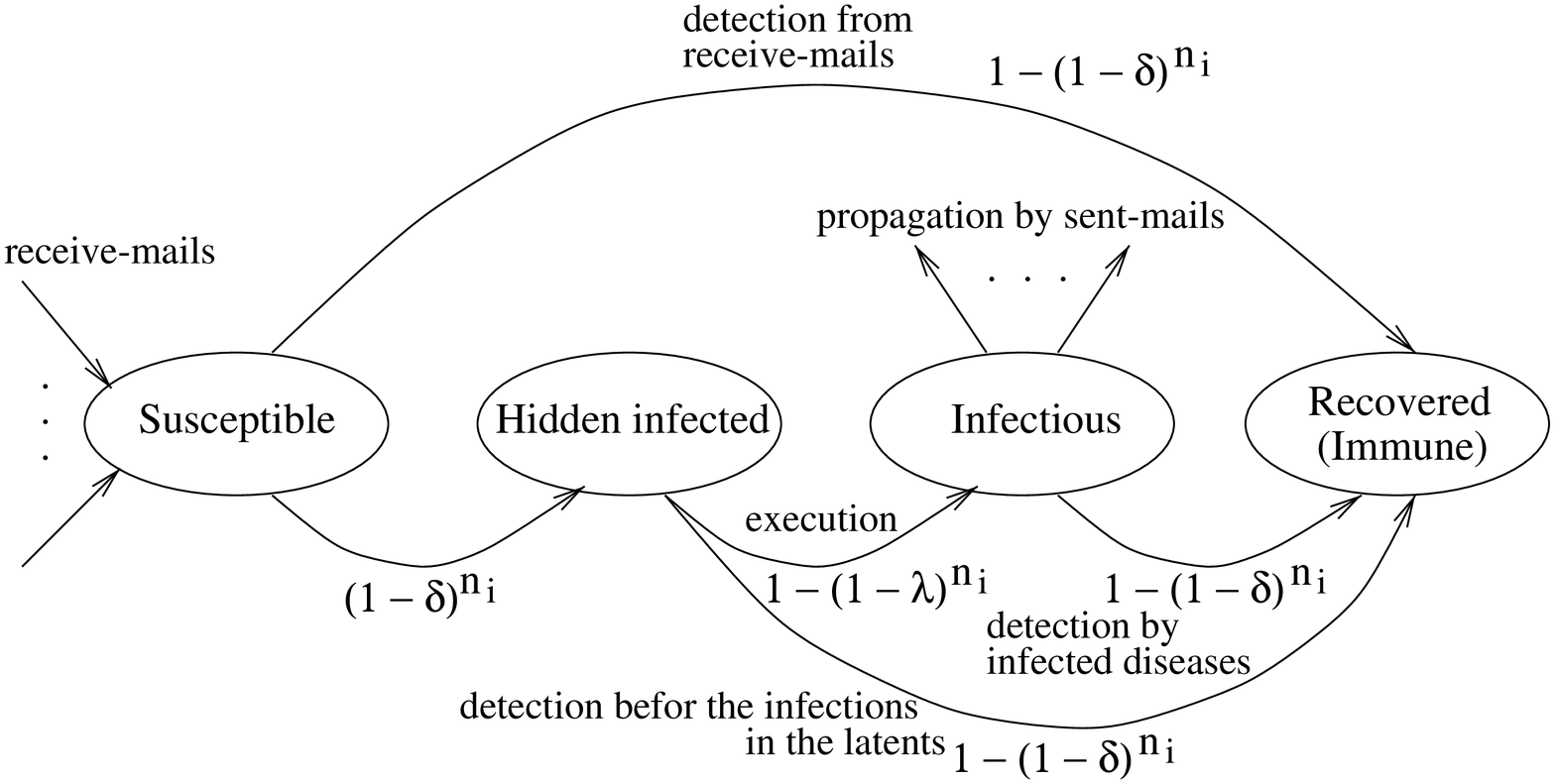}
\caption{S-H-I-R state transition diagram.} 
\label{fig_SHIR}
\end{figure}

We consider a network whose vertices and edges 
represent computers 
and the communication via e-mails between users.
The state at each computer $i = 1, \ldots, N$
is changed from the 
susceptible, hidden, infectious, and to the recovered by the remove of
viruses and installation of anti-viruses.
We make a realistic model in stochastic state transitions 
with probabilities of
the execution and the detection of viruses.
Fig. \ref{fig_SHIR} shows the state transitions,
where $\lambda$ and $\delta$
denote the execution rate from the hidden to 
the infectious
state and the detection rate from the special subjects or 
doubtful attachment files.
The probability at least one detection 
from the $n_{i}$ viruses on the computer is 
$1 - (1 - \delta)^{n_{i}}$, 
and the probability at least one execution is 
$1 - (1 - \lambda)^{n_{i}}$.
We assume the infected mail is not sent again for the same
communication partner
(sent it at only one time) to be difficult for the detection.
Thus, $n_{i}$ is at most the number of in-degree at each vertex.
In the stochastic SHIR model, the final state is the recovered or
immune by anti-viruses, if at least one infected mail is received.

\begin{figure}
  \begin{minipage}[htb]{.47\textwidth}
    \includegraphics[height=55mm]{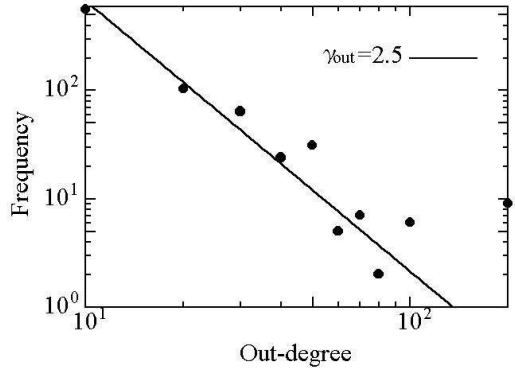} 
    \begin{center} (a) \end{center}
  \end{minipage} 
  \hspace{-5mm}
  \begin{minipage}[htb]{.47\textwidth}
    \includegraphics[height=55mm]{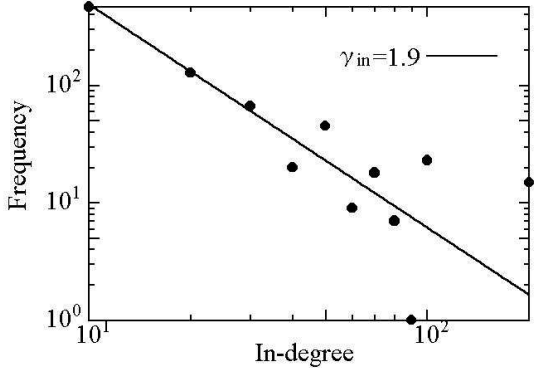} 
    \begin{center} (b) \end{center}
  \end{minipage} 
  \caption{Power-law degree distributions with the exponents 
$\gamma_{out} = 2.5$ and 
 $\gamma_{in} = 1.9$ for (a) sent-mails and (b) receive-mails 
 between users including the internal and the external \cite{Mikami01} .
 The frequency at degree $k$ is counted in the interval between 
 $[ k, \; k+10 ]$,
 except of the outer of more than 100 degree at $k = 200$.} 
  \label{fig_power_out_in}
\end{figure}

\subsection{The scale-free structure}
\begin{figure}
  \begin{minipage}[htb]{.47\textwidth}
    \includegraphics[height=45mm]{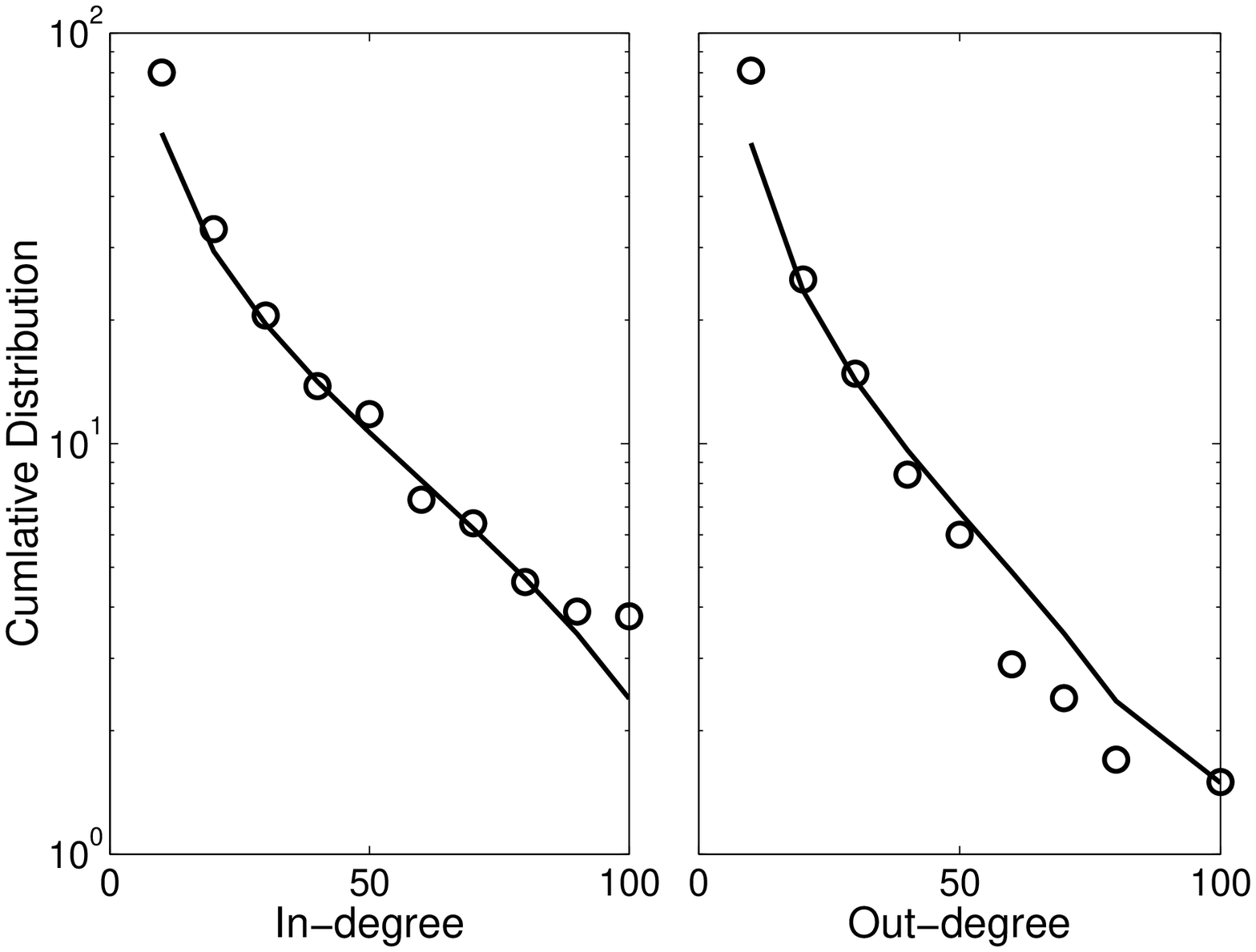} 
    \begin{center} (a) \end{center}
  \end{minipage} 
  \hfill 
  \begin{minipage}[htb]{.47\textwidth}
    \includegraphics[height=45mm]{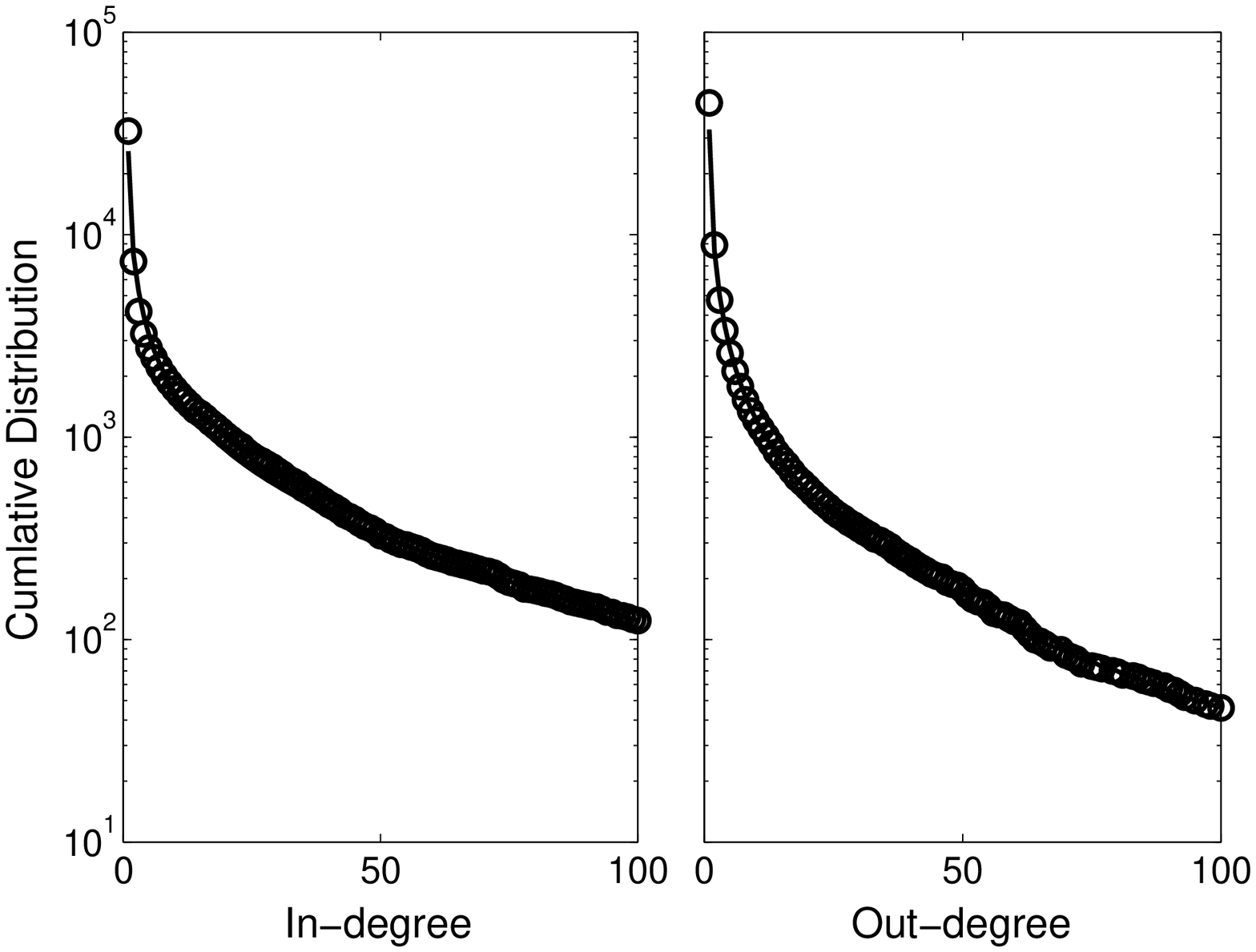} 
    \begin{center} (b) \end{center}
  \end{minipage} 
  \begin{minipage}[htb]{.47\textwidth} \vspace{5mm}
    \includegraphics[height=45mm]{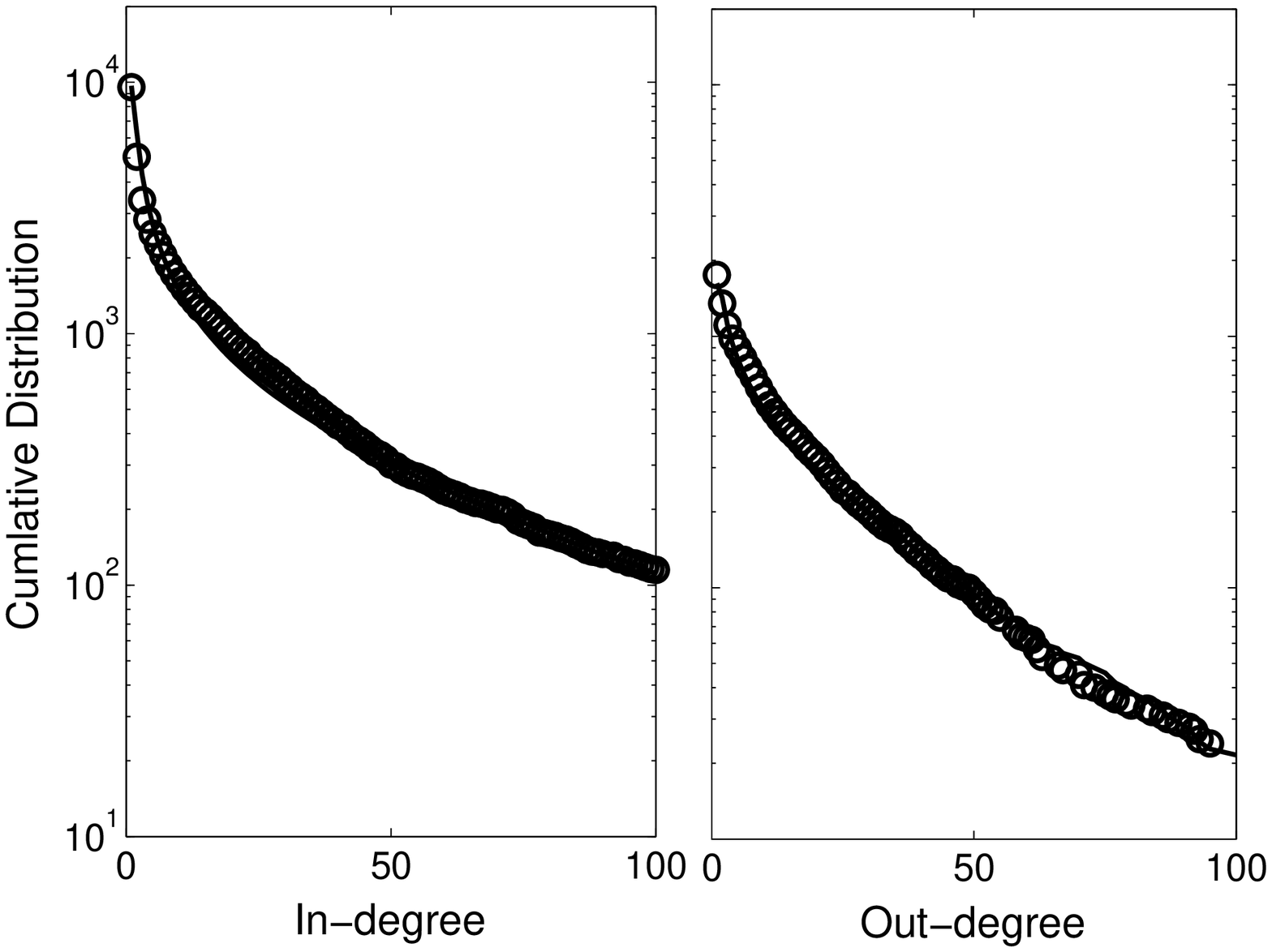} 
    \begin{center} (c) \end{center}
  \end{minipage} 
  \hfill 
  \begin{minipage}[htb]{.47\textwidth} \vspace{5mm}
    \includegraphics[height=45mm]{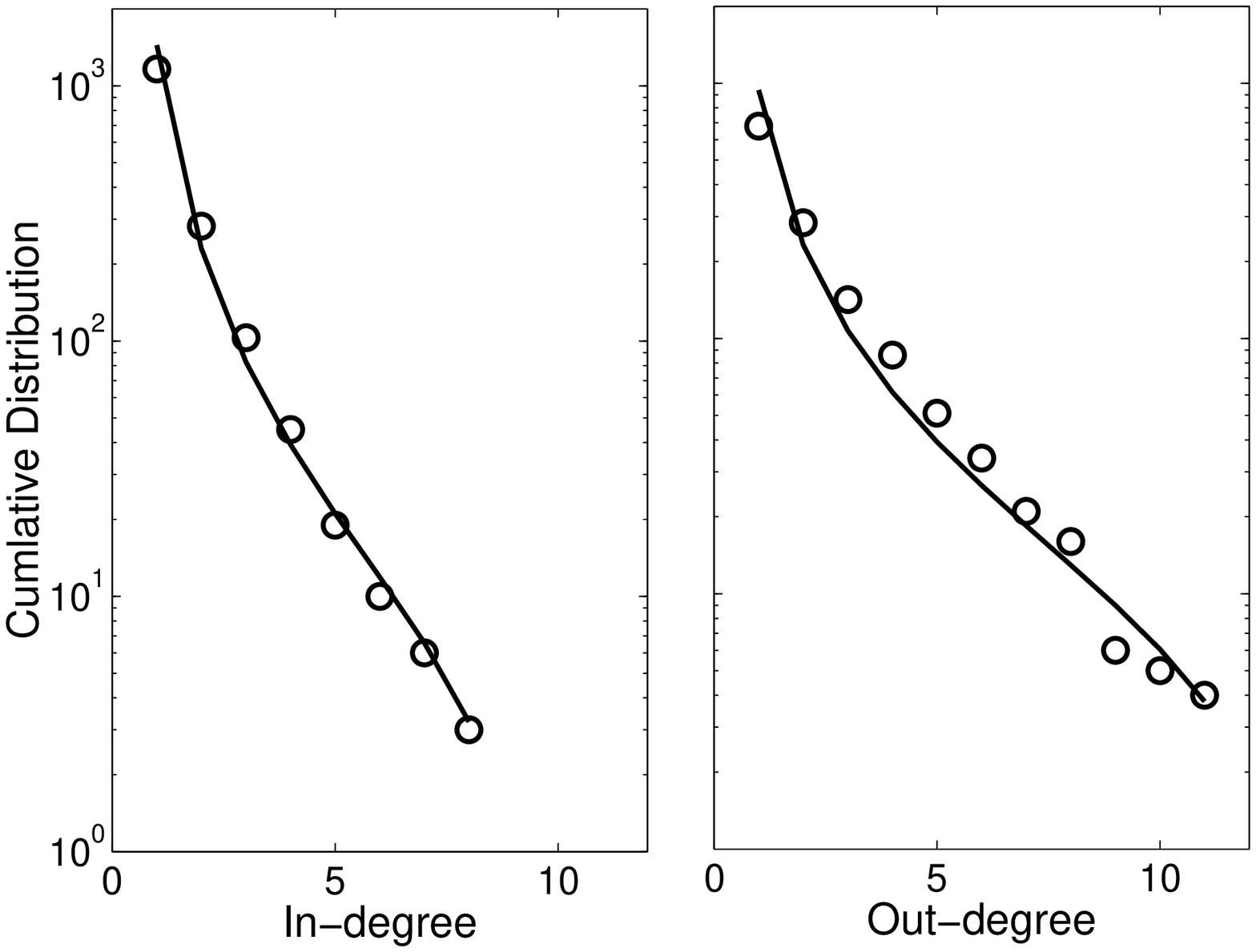} 
    \begin{center} (d) \end{center}
  \end{minipage} 
  \caption{The cumulative distributions of the in-degree and  
 out-degree for e-mail networks. (a) in the questionnaires 
 \cite{Mikami01}, and (b)-(d) in the server log files \cite{Data} for 
 (b) all of the internal and external nodes as e-mail users, 
 (c) the internal nodes including links from/to the external nodes, and 
 (d) the only internal networks.} 
  \label{fig_cum_dist}
\end{figure}

We show the network structure 
based on real data measured by questionnaires for 2,555 users
in a part of World Internet Project 2000 \cite{Mikami01}.
The distributions of both sent- and receive-mails follow a power-law
in Fig. \ref{fig_power_out_in},
the parameters are estimated as 
$\gamma_{out} = 2.5$, $\gamma_{in} = 1.9$, 
and the average number of mails par day ${\bar k} = 5 \sim 20$.
These values are close to the exponents 
$\gamma_{out} = 2.03 \pm 0.12$ and 
$\gamma_{in} = 1.49 \pm 0.12$ \cite{Ebel02}
estimated for the server log files of e-mails \cite{Data}.
In addition, the cumulative histograms of less than degree $k$ in 
Fig. \ref{fig_cum_dist} (a) have similar shapes to them in a larger
network of e-mail address books \cite{Newman02b}.
The solid lines in Fig. \ref{fig_cum_dist} 
correspond to non-cumulative distributions of 
the in-degree and out-degree estimated as 
stretched exponential 
\begin{description}
\item[(a) ] 
$P_{in}(k)  \sim k^{-1.97} \times \exp( -72.26 \times k^{-316.23})$,
$P_{out}(k) \sim k^{-2.29} \times \exp( -56.98 \times k^{-247.39})$,
\item[(b) ] 
$P_{in}(k)  \sim k^{-1.82} \times \exp( 0.63 \times k^{-52.99})$,
$P_{out}(k) \sim k^{-2.02} \times \exp( 0.6 \times k^{-59.23})$,
\item[(c) ] 
$P_{in}(k)  \sim k^{-1.75} \times \exp( -0.71 \times k^{-53.42})$,
$P_{out}(k) \sim k^{-1.39} \times \exp( -1.43 \times k^{-137.95})$,
\item[(d) ] 
$P_{in}(k)  \sim k^{-2.87} \times \exp( 3.54 \times k^{-0.048})$,
$P_{out}(k) \sim k^{-2.49} \times \exp( 3.23 \times k^{-0.002})$.
\end{description}
In all of them, the factor of power law as a scale-free network is
dominant.
Note that the in-degree distribution in Fig. \ref{fig_cum_dist} (d) 
is the most close to the exponential distribution
in \cite{Newman02b} with a strong cut-off,
and that both data consists of only the internal networks.
However, as in \cite{Davidsen02} \cite{Newman02b},
we must further discuss about the reason why exponential
in-degree distribution appears in only the internal networks.
This is beyond the scope of this paper.
The non-exponential distributions may be caused by the limited size of the
sample, or by that the eliminated links from the external nodes have an 
impact on the generation of hubs in a scale-free network.

\subsection{The $(\alpha, \; \beta)$ model}
With the estimated parameters, 
we generate a SF network for the contact
relations between e-mail users, by applying the  simple 
$(\alpha, \; \beta)$ model \cite{Kumar99}, in which 
the slopes of power-law 
$\gamma_{in} \approx \frac{1}{1 - \alpha}$ and 
$\gamma_{out} \approx \frac{1}{1 - \beta}$ 
are 
controlled by the $\alpha$-$\beta$ coin in 
Table \ref{table_alpha_beta} (in the case of e-mails 
$\alpha = 0.4736$ and $\beta = 0.6$).
Growing with a new vertex at each step, 
$k$ edges are added as follows.
As the terminal, the coin chooses a new vertex with probability 
$\alpha$ and 
an old vertex with probability $1 - \alpha$
in proportion to its in-degree.
As the origin, the coin chooses a new vertex with probability 
$\beta$ and an old vertex with probability $1- \beta$
in proportion to its out-degree.
According to both the growth and the preferential attachment
\cite{Albert01}\cite{Barabasi99}, 
the generation processes are repeated until the required size $N$ 
is obtained as 
a connected component without self-loops and multi-edges.
The $(\alpha, \; \beta)$ model generates both of edges from/to a new
vertex and edges between old vertices, the processes are somewhat
analogous to ones 
in the generalized BA model \cite{Albert00b}\cite{Albert01}.

\begin{table}[htb]
\begin{center}
\begin{footnotesize}
\begin{tabular}{ccc} \hline\hline
probability     & $\alpha$ & $1 - \alpha$ \\ \hline
$\beta$         & self-loop at new vertex & origin: new, terminal: old\\ \hline
$1 - \beta$     & terminal: new, origin: old & both of old vertices\\ \hline\hline
\end{tabular}
\end{footnotesize}
\vspace{2mm}
\caption{Directed edge generation by the $\alpha$-$\beta$ coin.}
\label{table_alpha_beta}
\end{center}
\end{table}

\section{SIMULATIONS FOR STOCHASTIC MODEL}
\begin{figure}
  \begin{minipage}[htb]{.47\textwidth}
    \includegraphics[width=70mm]{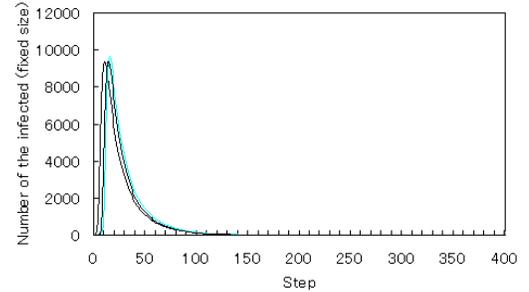} 
    \begin{center} (a) \end{center}
  \end{minipage} 
  \hfill 
  \begin{minipage}[htb]{.47\textwidth}
    \includegraphics[width=70mm]{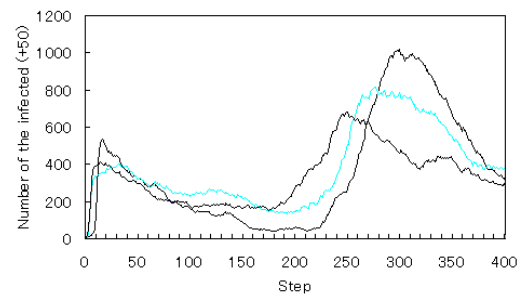} 
    \begin{center} (b) \end{center}
  \end{minipage} 
  \begin{minipage}[htb]{.47\textwidth}
    \includegraphics[width=70mm]{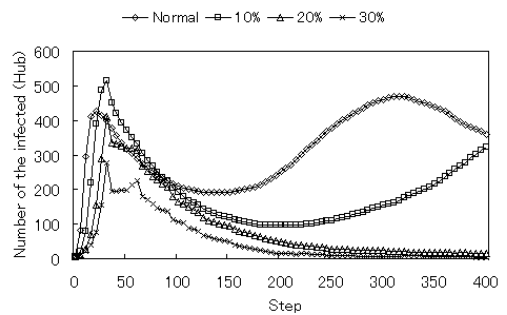} 
    \begin{center} (c) \end{center}
  \end{minipage} 
  \hfill 
  \begin{minipage}[htb]{.47\textwidth}
    \includegraphics[width=70mm]{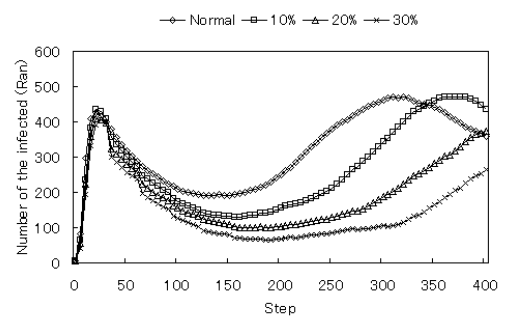} 
    \begin{center} (d) \end{center}
  \end{minipage} 
  \caption{Typical behavior of the spread on SF networks 
	in (a) a closed system and (b) an open system with
        simultaneously progress of both spread of viruses and growth
        of network.
	The lines show the differences in stochastic
	state transitions.
	The effects of immunization are shown as the averages in the
	open system for 
	(c) hub and 
	(d) random immunization.
	The open diamond, square, triangle, 
	and cross marks are corresponding to 
    	the normal detection by the state transitions, 
	immunization of the 
	10 \%, 20 \%, and 30 \%, respectively.}
  \label{fig_I_HubRan}
\end{figure}

We 
study the typical behavior in the SHIR model on the SF networks.
In the following simulations, we set the execution rate 
$\lambda = 0.1$, the detection rate $\delta = 0.04$, 
the average number of edges ${\bar k } = 6.6$, 
and initial infection sources of randomly chosen five vertices
(the following results are similar to other small values
$\lambda = 0.2, 0.3$ and $\delta = 0.05, 0.06$).
These small values are realistic,
because computer viruses are not recognized before the prevalence 
and it may be executed by some users.
We note the parameters are related to the sharpness of 
increasing/decreasing infections up/down ($\delta$ is more sensitive).
It is well known, in a closed system of the SHIR model, 
the number of infected computers (the hidden and infectious states)
is initially increased and saturated, finally converged to zero as the
extinction.
While the pattern may be different in an open system, indeed, 
oscillations have been described
by a deterministic Kermack-McKendrik model \cite{Shigesada97}.
However a constant population (equal rates of the birth and the death) or 
territorial competition has been mainly discussed in the model,
the growth of computer network is obviously more rapid, 
and the communications in mailing are not competitive.
Thus, we consider a growing system, in which 
$50$ vertices and the corresponded new $\bar k$ edges  
are added at every step, 
from an initial SF network with $N = 400$ 
up to 20350 at 400 steps.
Here, one step is corresponding to a day (400 steps $\approx$ a
year).
These values of $\lambda$, $\delta$, $\bar{k}$, and the growth rate 
are only examples with something of reality for simulations, since 
the actual values depends on the observed period are still unknown.
As shown in Fig. \ref{fig_I_HubRan}(a)(b), 
the phenomena of persistent recoverable prevalence are found in the open
system, but not in the closed system.

\begin{figure}
  \begin{minipage}[htb]{.47\textwidth}
    \includegraphics[width=70mm]{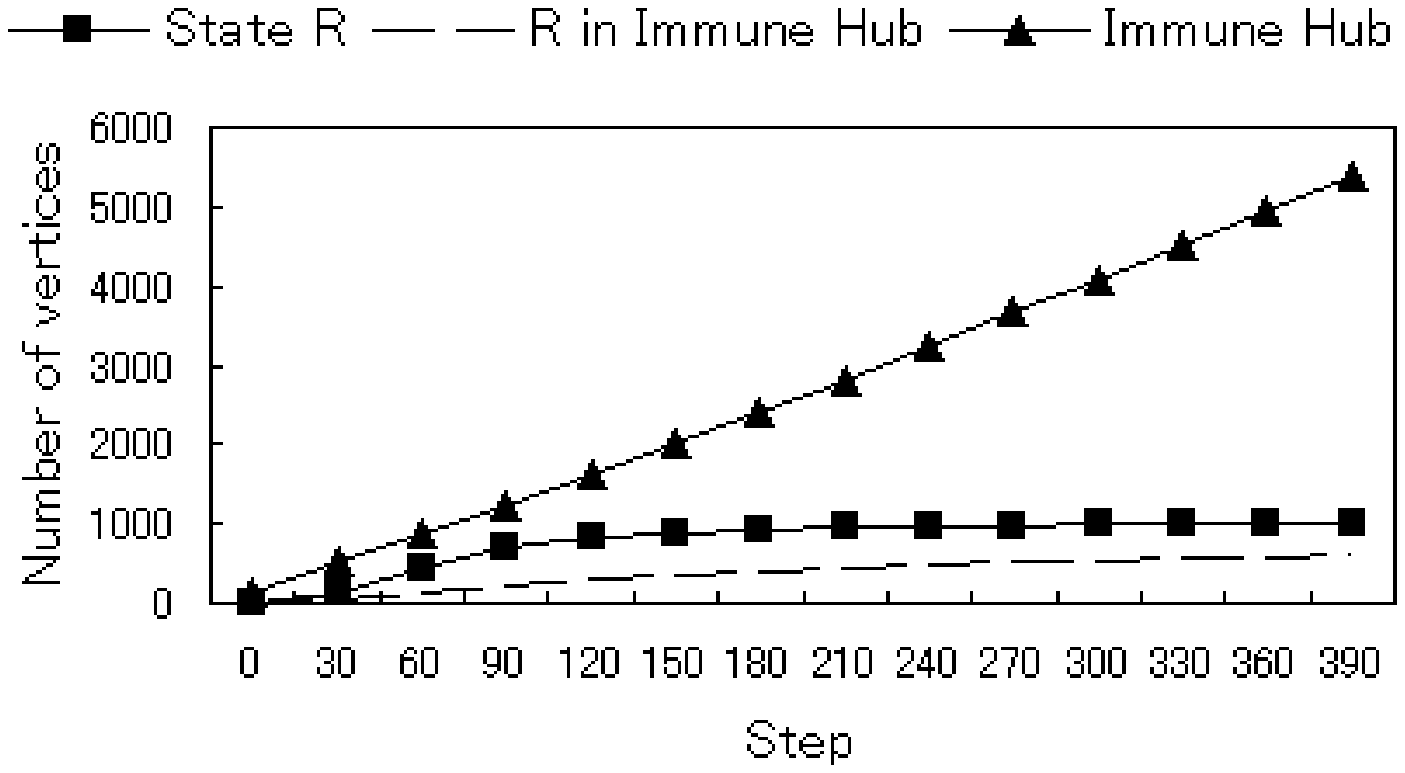} 
    \begin{center} (a) \end{center}
  \end{minipage} 
  \hfill 
  \begin{minipage}[htb]{.47\textwidth}
    \includegraphics[width=70mm]{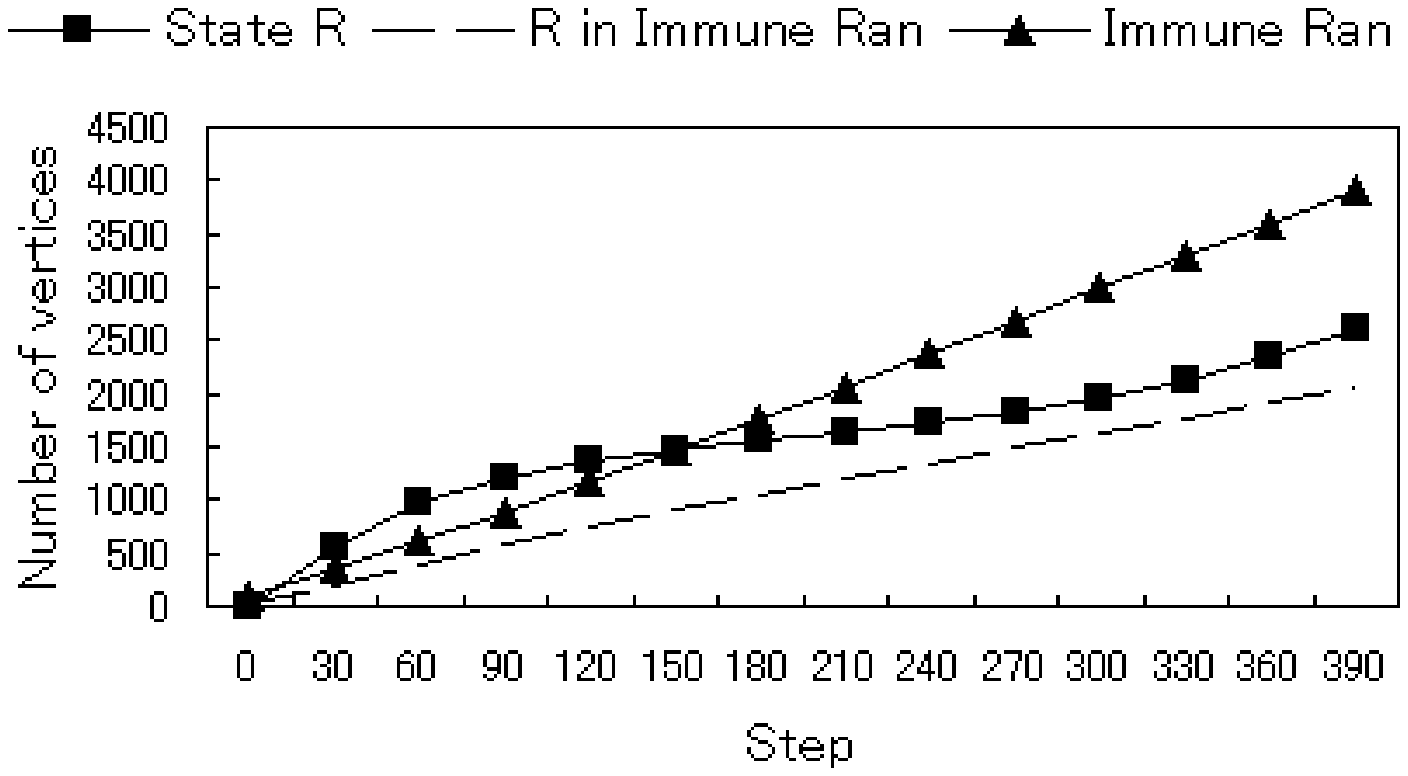} 
    \begin{center} (b) \end{center}
  \end{minipage} 
  \caption{Number of vertices in the recovered state 
	by 
	(a) hub and 
	(b) random immunization of the 30 \%. 
	Each of them is the average value for recoverable prevalence 
    	in 100 trials. The dashed lines represent the number of
    	vertices that are already changed to the recovered states before the 
    	immunization.}
  \label{fig_stateR_lin}
\end{figure}

To prevent the wide spread of infections, we investigate how to assign 
anti-virus softwares onto the SF networks.
We verify the effectiveness of the targeted immunization for hubs
even in the cases of recoverable prevalence.
Fig. \ref{fig_I_HubRan}(c)(d) show
the average number of infected computers with recoverable prevalence 
in 100 trials,
where immunized vertices are randomly selected or as hubs according to the
out-degree order of the 10 \%, 20 \%, 30 \% of growing size at every 30
steps (corresponded to a month).
The number is decreased as larger immune rates for hubs,
viruses are nearly extinct (there exists only few viruses) 
in the 30 \% 
as marked by $\times$ in Fig. \ref{fig_I_HubRan}(c).
While it is also decreased as larger immune rates for randomly
selected vertices,
however 
they are not extinct even in the 30 \%  
as marked by $\times$ in Fig. \ref{fig_I_HubRan}(d).
Fig. \ref{fig_stateR_lin}(a)(b) show the number of recovered
states by the hub and random immunization of 
the 30 \% (triangle marks) for the comparison with 
the normal detections (rectangle marks).
The immunized hubs are dominant than the normal detections 
in Fig. \ref{fig_stateR_lin}(a).
However, there is no such difference for the random immunization 
in Fig. \ref{fig_stateR_lin}(b).
In the case of the 10 \%, the relation is exchanged; 
the number of detections is larger than that of both hub and
random immunization. It is the intermediate in the case of the 20 \%. 
From these results, we remark 
the targeted immunization for hubs strongly prevents  
the spread of infections in spite of 
the totally fewer recovered states 
than that in random immunization.

\section{ANALYSIS FOR DETERMINISTIC MODEL}
Although 
the stochastic SHIR model is realistic, the analysis is
very difficult in the open system.
Thus, we analyze simpler deterministic SIR models for the spreading of
computer viruses  
to understand the mechanisms of recoverable prevalence and 
extinction by the immunization.
We consider the time evolutions of $S(t) > 0$ and $I(t) > 0$ ($t \geq 0$),
which are the number of susceptible and infected vertices.
We assume that infection sources exist in an initial network, 
and that both network growth and the spread of viruses 
are progressed in continuous time as an approximation.
In addition, we have no specific rules in growing, but consider 
a linearly growing network size and 
the distribution of connections on an undirected connected graph 
as a consequence.

\subsection{Homogeneous SIR model}
\begin{figure}
  \begin{minipage}[htb]{.47\textwidth}
    \includegraphics[width=60mm]{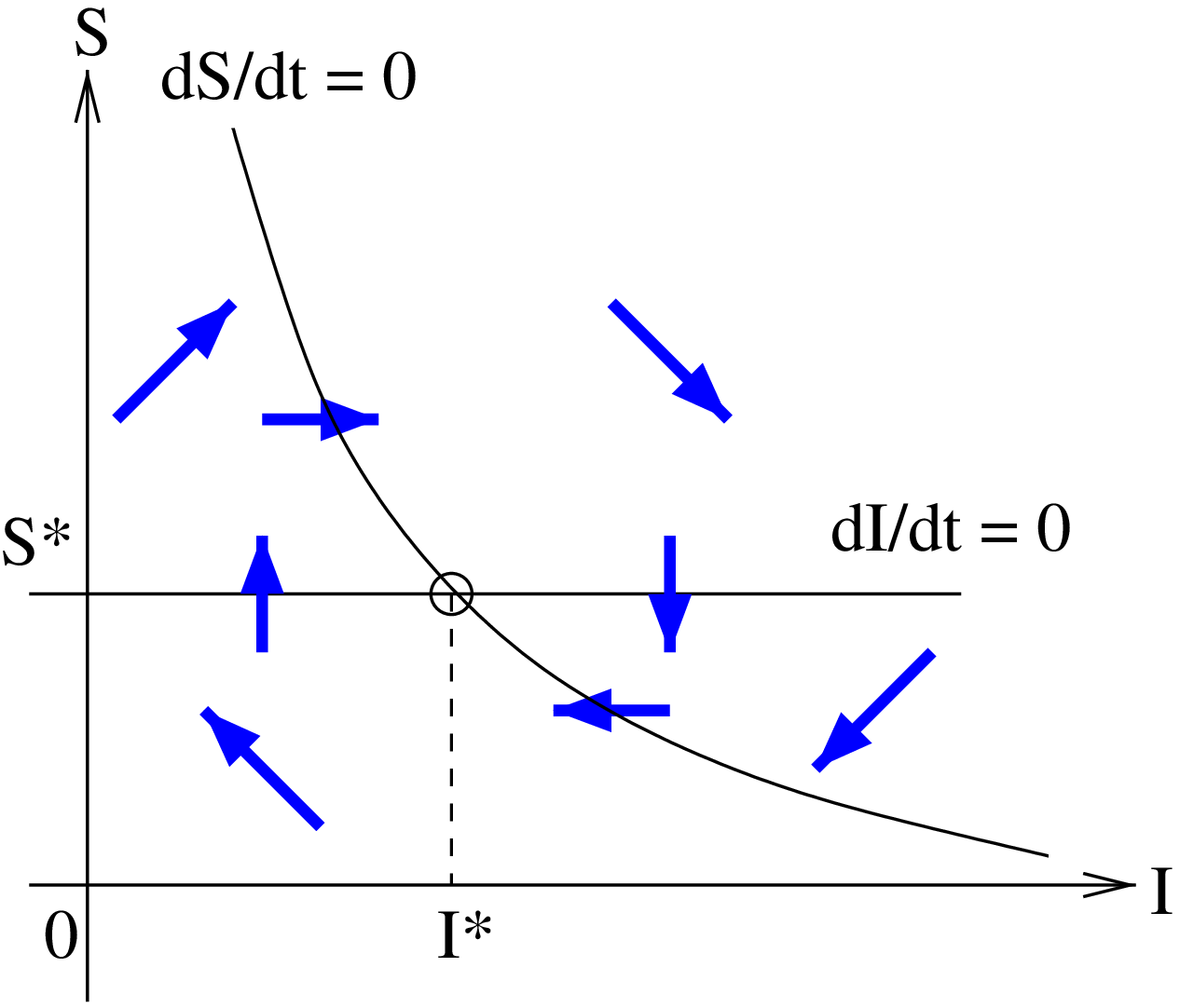} 
    \begin{center} (a) \end{center}
  \end{minipage} 
  \hfill 
  \begin{minipage}[htb]{.47\textwidth}
    \includegraphics[width=60mm]{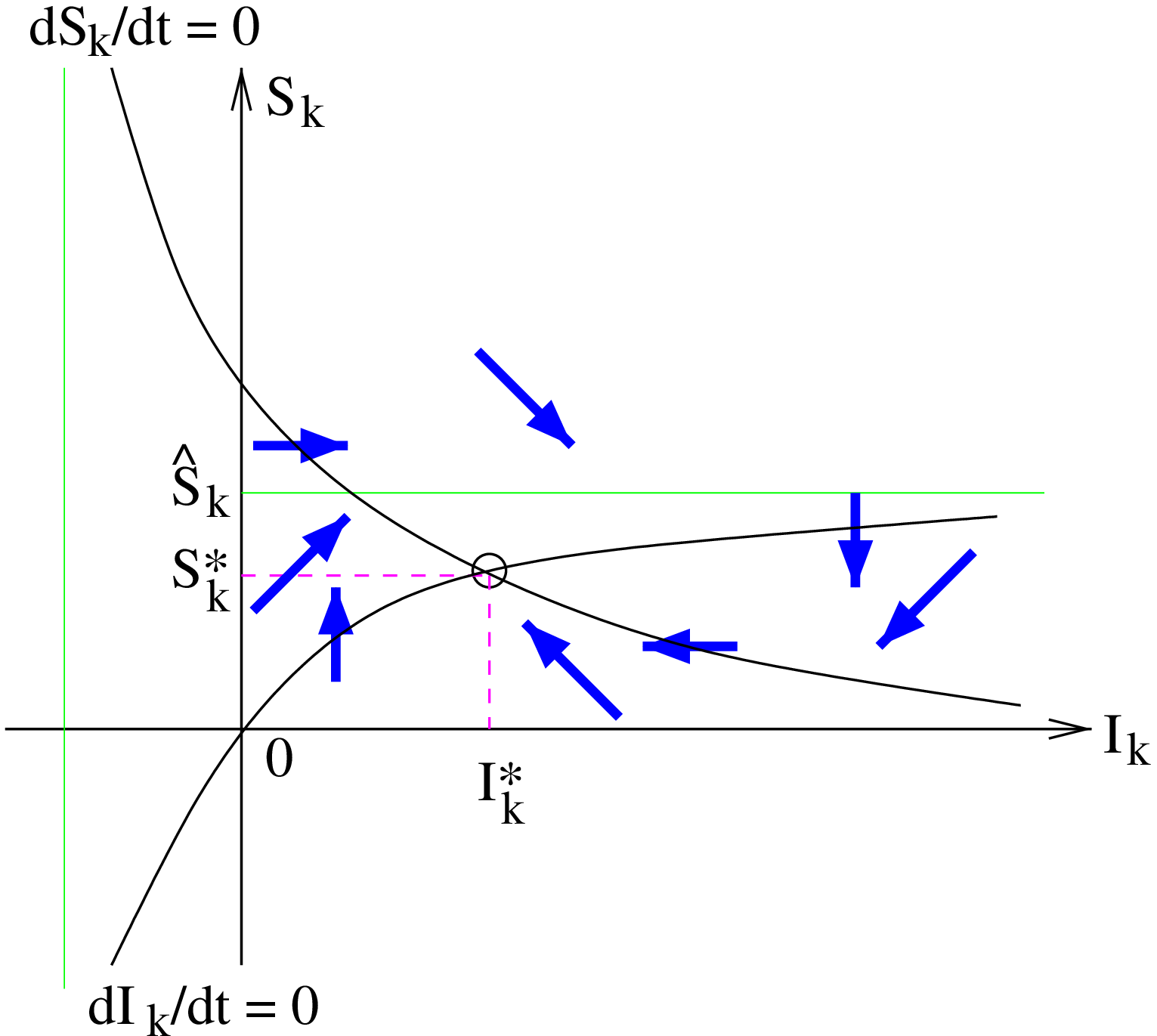} 
    \begin{center} (b) \end{center}
  \end{minipage} 
  \caption{Nullclines and the vector fields for 
	(a) homogeneous and 
	(b) heterogeneous SIR models.
	The state in both cases is converged to an equilibrium point
	with a damped oscillation, which is corresponded to persistent
        recoverable prevalence around the non-zero level 
        $I^{*}$ or $I^{*}_{k}$.}
  \label{fig_vector_SIR}
\end{figure}

As the most simple case, 
in the homogeneous networks with only the detection of viruses, 
the time evolutions are given by 
\begin{eqnarray}
  \frac{d S(t)}{dt} & = & -b <k> S(t) I(t) + a,
	\label{eq_homo_S_recover}\\
  \frac{d I(t)}{dt} & = & -\delta_{0} I(t) + b <k> S(t) I(t), 
	\label{eq_homo_I_recover}
\end{eqnarray}
where $a > 0$ and $0 < b, \; \delta_{0} < 1$ denote the growth,
infection, and detection rates, respectively.
$<k> \stackrel{\rm def}{=} \sum_{k} k P(k)$ 
is the average number of connections with a probability $P(k)$.
The term $S(t)I(t)$ represents the frequency of contact relations.
Note that the number of recovered vertices $R(t)$ is 
a shadow variable defined by 
$\frac{d R(t)}{d t} = \delta_{0} I(t)$.
From the network size $N(t) = S(t) + I(t) + R(t)$, 
the solution is given by $N(t) = N(0) + a t$ as a linear growth.
Fig. \ref{fig_vector_SIR}(a) shows the nullclines of 
\[
  \frac{d S}{dt} = 0: \; S = \frac{a}{b <k> I},
\]
\[
  \frac{d I}{dt} = 0: \; 
	S = S^{*} \stackrel{\rm def}{=} \frac{\delta_{0}}{b <k>}, \;
  (I \neq 0)
\]
for Eqs. (\ref{eq_homo_S_recover})(\ref{eq_homo_I_recover}).
The directions of vector filed are 
defined by the positive or negative signs of
$\frac{d S}{dt}$ and $\frac{d I}{dt}$.
There exists a stable equilibrium point $(I^{*}, S^{*})$,
$I^{*}  \stackrel{\rm def}{=} \frac{a}{\delta_{0}}$.
The states of $S$ and $I$ are converged to the point 
with a damped oscillation.
We can easily check the real parts of eigenvalues for the Jacobian 
are negative at the point.

\subsection{Heterogeneous SIR model}
Next, we 
consider the heterogeneous SF networks 
at the mean-field level, in which the connectivity correlations are
neglected \cite{Moreno02}.
We know that static and grown 
networks have different properties for the
size of giant component \cite{Dorogovtsev01}
and the connectivity correlations \cite{Callaway01}\cite{Krapivsky01}
even if the degree distributions are the same.
In particular, the correlations may have influence on the spread, 
however they are not found 
in all growing network models or real systems.
We have experientially observed the correlations  
are very week in the 
($\alpha$, $\beta$) model in the previous simulations 
as similar to the nearest neighbors 
average connectivity 
in the generalized BA model rather
than the fitness model or AS in the Internet \cite{Satorras01b}.
At least, non-correlation seems to be not crucial 
for the absence of epidemic threshold 
\cite{Dezso02}\cite{Moreno02}\cite{Satorras01a}\cite{Satorras02},
the existence of correlations 
is much still less nontrivial in e-mail networks.
Although the mean-field approach by neglecting the correlations 
in macroscopic equations at a large network size
is a crude approximation method, 
it is useful for understanding the mechanisms of the spread in
growing networks, as far as it is qualitatively similar to 
the behavior of viruses 
in the stochastic model or observed real data.
Indeed, the following results are consistent with the analysis for 
correlated cases \cite{Hayashi03}, 
except of the quantitative differences.

We introduce a linear kernel  
\cite{Krapivsky01} as $N_{k}(t) \sim a_{k} \times t$, 
$N_{k}(t) = S_{k}(t) + I_{k}(t) + R_{k}(t)$, which are sum of the
numbers of susceptible, infected, and recovered vertices with 
connectivity $k$, 
and the growth rate  
$a_{k} \stackrel{\rm def}{=} A k^{-\nu}$, $A > 0$, $\nu > 2$. 
Note that the total  
$N(t) = \sum_{k} N_{k}(t) \sim \left( \sum_{k} a_{k} \right) \times t$ 
means a linear growth of network size. 
Since the maximum degree increases as progressing the time 
and approaches to infinity, 
it has a nearly constant growth rate  
$\sum_{k = m}^{\infty} a_{k} \sim \int_{m}^{\infty} A k^{- \nu} d k 
= \frac{A m^{1 - \nu}}{\nu - 1}$
for large $t$.
As shown in \cite{Krapivsky01}, the introduction of linear kernel 
is not contradiction with the 
preferential (linear) attachment \cite{Albert01} \cite{Barabasi99}.

At the mean-field level in a somewhat large network
with only the detection of viruses, 
the time evolutions of $S_{k} > 0$ and $I_{k} > 0$ are given by 
\begin{eqnarray}
  \frac{d S_{k}(t)}{dt} & = & 
	-b k S_{k}(t) \Theta(t) + a_{k}, 
	\label{eq_hete_S_recover} \\
  \frac{d I_{k}(t)}{dt} & = & 
	-\delta_{0} I_{k}(t) + b k S_{k}(t) \Theta(t),
	\label{eq_hete_I_recover}
\end{eqnarray}
where the shadow variable $R_{k}(t)$ is implicitly defined 
by $\frac{d R_{k}(t)}{d t} = \delta_{0} I_{k}(t)$.
The factor 
$\Theta(t) \stackrel{\rm def}{=} \sum_{k} c_{k} I_{k}(t)$,
$c_{k} \stackrel{\rm def}{=} \frac{k P(k)}{<k>}$, 
represents the expectation that any given edge points to an infected
vertex.

\begin{figure}
  \begin{minipage}[htb]{.47\textwidth}
    \includegraphics[width=30mm]{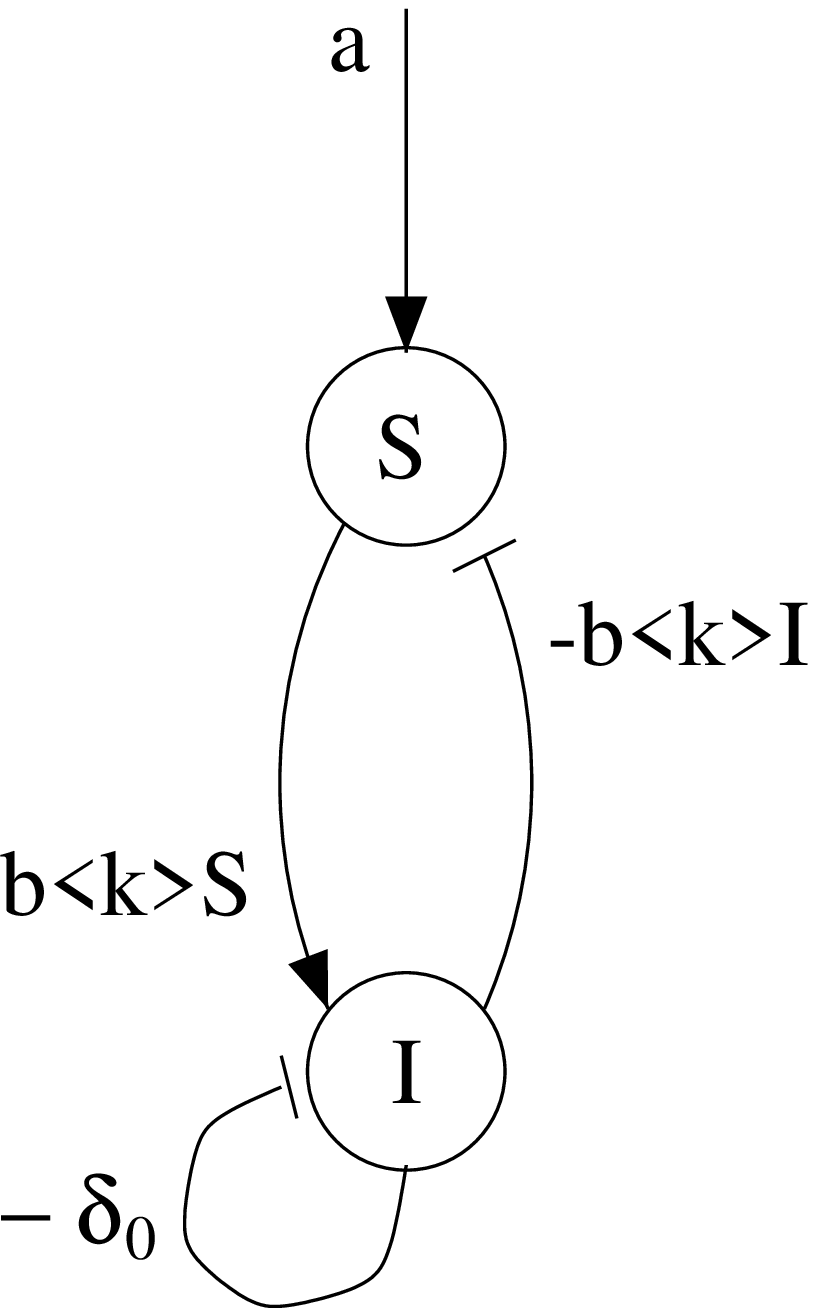} 
    \begin{center} (a) \end{center}
  \end{minipage} 
  \hfill 
  \begin{minipage}[htb]{.47\textwidth}
    \includegraphics[width=60mm]{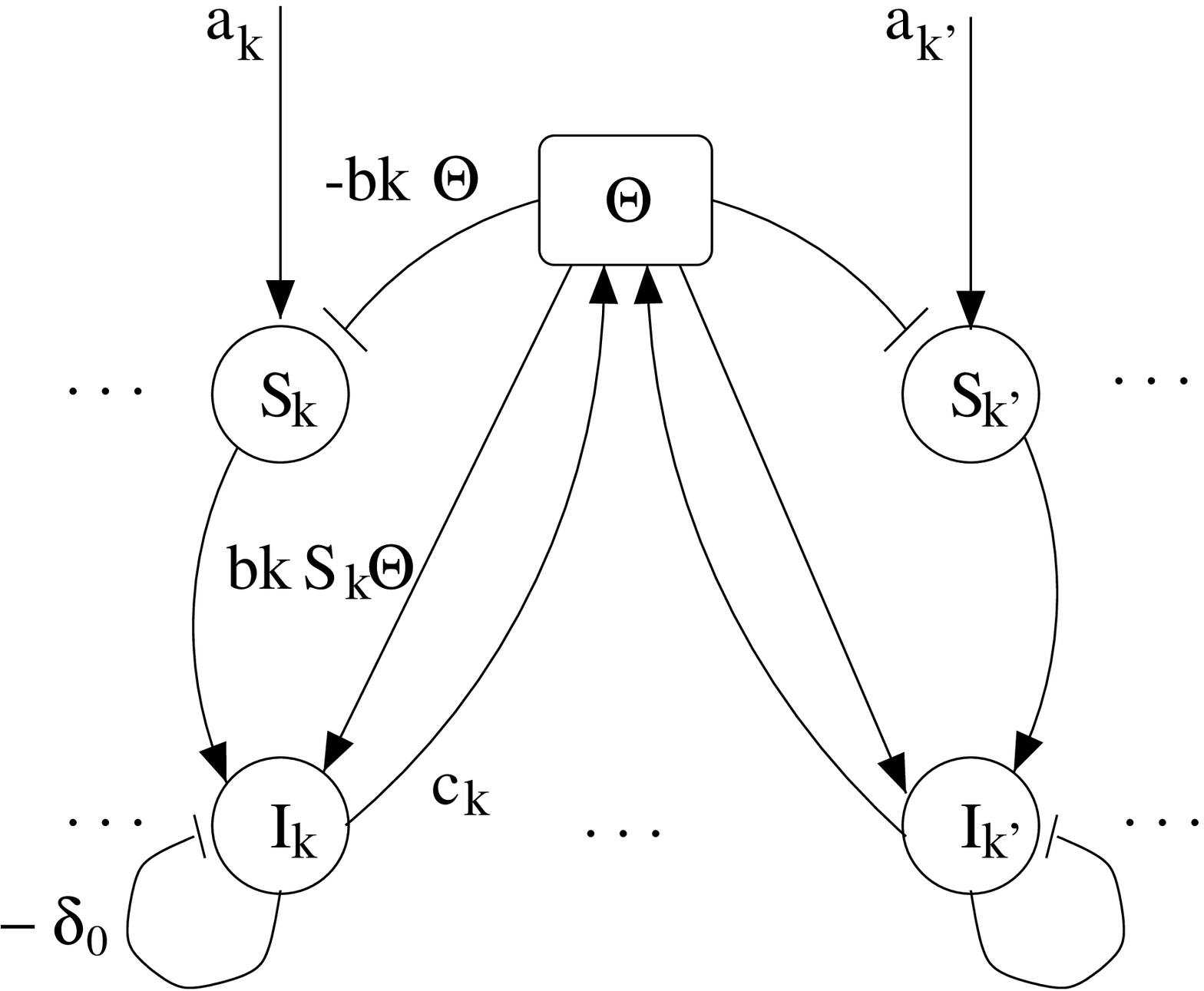} 
    \begin{center} (b) \end{center}
  \end{minipage} 
  \caption{Oscillators for 
	(a) homogeneous and 
	(b) heterogeneous SIR models in the open system.
	They consist of S-I pairs with 
	excitatory: $\rightarrow$
	and inhibitory: $\dashv$ connections, 
	and an input bias $a$ or $a_{k}$ of the growth rate.
	The factor $\Theta$ acts as a global inhibition or
    	excitation.}
  \label{fig_oscillator}
\end{figure}

We consider a section of $I_{k'} = I^{*}_{k'}$: const. 
for all $k' \neq k$.
Fig. \ref{fig_vector_SIR}(b) shows the nullclines of 
\[
  \frac{d S_{k}}{dt} = 0: \;
	S_{k} = \frac{a_{k}}{kb\Theta} = \frac{a_{k}}{k b c_{k} I_{k} + k b
	\sum_{k'} c_{k'} I^{*}_{k'}}, 
\]
\[
  \frac{d I_{k}}{dt} = 0: \;
	S_{k} = \frac{\delta_{0} I_{k}}{kb\Theta} 
	= \frac{\delta_{0} I_{k}}{k b c_{k} I_{k} + k b
	\sum_{k'} c_{k'} I^{*}_{k'}}, 
\]
and the vector filed for Eqs. 
(\ref{eq_hete_S_recover})(\ref{eq_hete_I_recover}).
There exists 
a stable equilibrium point $(I^{*}_{k}, S^{*}_{k})
	\stackrel{\rm def}{=} 
	(\frac{a_{k}}{\delta_{0}}, \frac{a_{k}}{k b \Theta^{*}})$,
because of 
\[
 \exists \Theta^{*} = \sum_{k \geq m} c_{k} I^{*}_{k} \sim
 \frac{A \gamma m^{-\gamma}}{\delta_{0}} \int_{m}^{\infty} 
 k^{-(\nu + \gamma + 1)} d k = 
 \frac{A \gamma m^{-\nu}}{\delta_{0} (\nu + \gamma)},
\]
by using 
$c_{k} = \gamma \times m^{\gamma} \times k^{-(\gamma + 1)}$
for the generalized BA model \cite{Moreno02} with a 
power-law degree distribution 
$P(k) = (1 + \gamma) m^{1 + \gamma} k^{-2 -\gamma}$,
$<k> = \frac{1 + \gamma}{\gamma} m$, 
(which includes the simple BA model \cite{Barabasi99} at $\gamma = 1$).
On these state spaces in Fig. \ref{fig_vector_SIR}(a)(b), 
only the case of $a = 0$ or $a_{k} = 0$ gives the
extinction: $I^{*} = 0$ or $I^{*}_{k} = 0$.
It means that 
we must stop the growing to prevent the infections 
by the detection. 
In addition, the homogeneous and heterogeneous systems are
regarded as oscillators in Fig. \ref{fig_oscillator}(a)(b).

\subsection{Effect of immunization}
We 
study the effect of random and hub immunization. 
With the randomly immune rate $0 < \delta_{r} < 1$,
the time evolutions are given by 
\begin{eqnarray}
  \frac{d S_{k}(t)}{dt} & = & 
	-b k S_{k}(t) \Theta(t) + a_{k} - \delta_{r} S_{k}(t),
	\label{eq_hete_S_immune}\\
  \frac{d I_{k}(t)}{dt} & = & 
	-\delta_{0} I_{k}(t) + b k S_{k}(t) \Theta(t)
	- \delta_{r} I_{k}(t),
	\label{eq_hete_I_immune}
\end{eqnarray}
where the shadow variable $R_{k}(t)$ is also defined by 
$\frac{d R_{k}(t)}{d t} 
	= \delta_{0} I_{k}(t) + \delta_{r} (S_{k}(t) + I_{k}(t))$.

We also consider a section of $I_{k'} = I^{*}_{k'}$: const. 
for all $k' \neq k$.
From the nullclines of Eqs. 
(\ref{eq_hete_S_immune}) and (\ref{eq_hete_I_immune})
with random immunization, 
there exists a stable equilibrium point $(I^{*}_{k}, S^{*}_{k})
	\stackrel{\rm def}{=} 
	( \frac{a_{k} - \delta_{r} S^{*}_{k}}{\delta_{0} + \delta_{r}}, 
	\frac{a_{k}}{\delta_{r} + k b \Theta^{*}} )$,
if the solution 
\[
 \Theta^{*} = \sum_{k} c_{k} I^{*}_{k} =
 \frac{1}{\delta_{0} + \delta_{r}} \sum_{k} a_{k} c_{k} 
 \left( 1 - \frac{\delta_{r}}{\delta_{r} + b k \Theta^{*}} \right)
 \stackrel{\rm def}{=} f(\Theta^{*}),
\]
is self-consistent at the point.
The condition is given by 
\[
 \begin{array}{lll}
 \frac{d f}{d \Theta} |_{\Theta = 0} & \approx & 
 \frac{A b}{\delta_{r} (\delta_{0} + \delta_{r})} \int_{m}^{\infty}
 \gamma m^{\gamma} k^{-(\gamma + \nu)} d k \\
 & = & 
  \frac{A b \gamma m^{-(\nu -1)}}{\delta_{r} (\delta_{0} + \delta_{r}) 
 (\gamma + \nu -1)} > 1.
 \end{array}
\]
In this case, the state space is the same as shown in
Fig. \ref{fig_vector_SIR}(b).

\begin{figure}
  \begin{minipage}[htb]{.47\textwidth}
    \includegraphics[width=60mm]{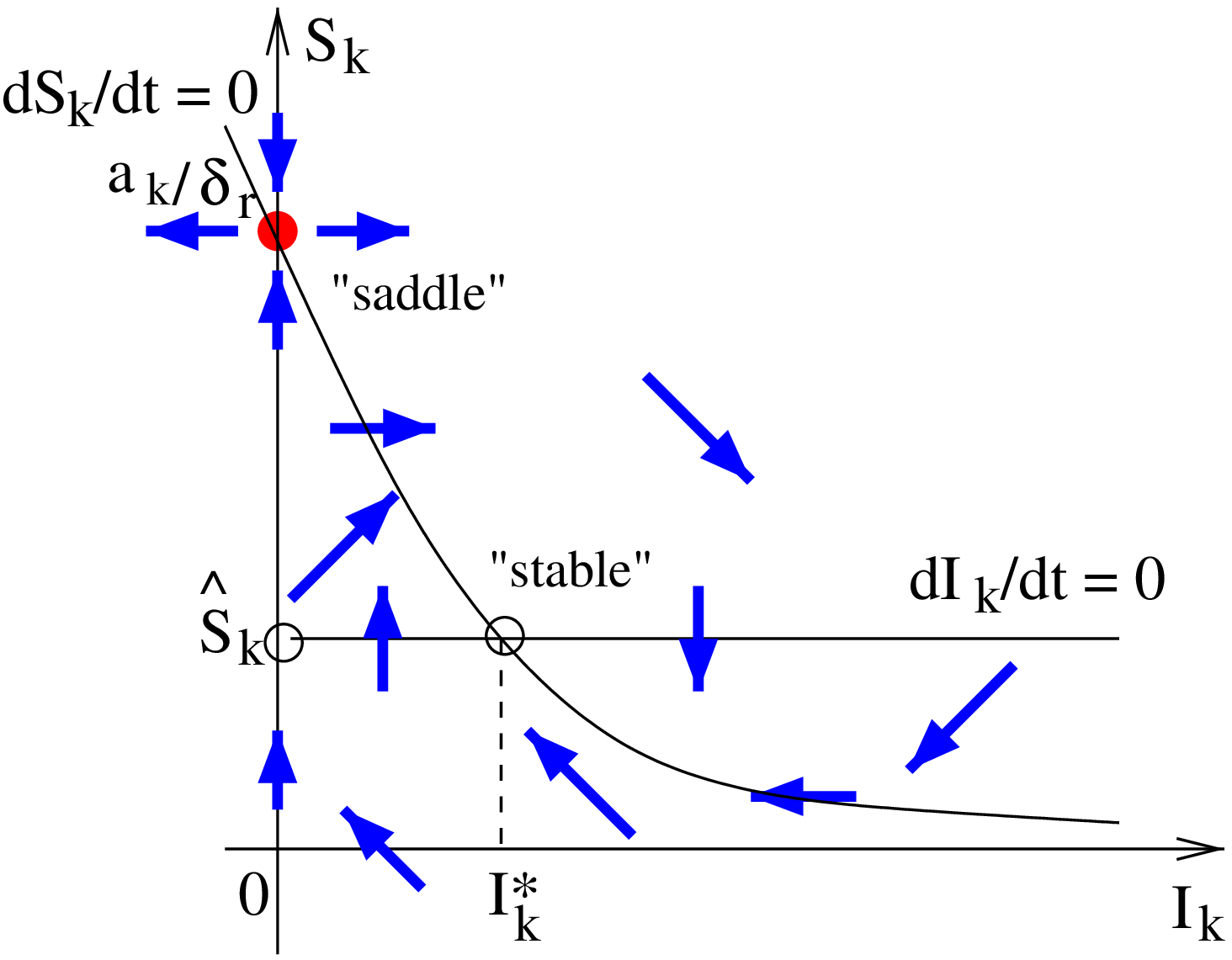} 
    \begin{center} (a) \end{center}
  \end{minipage} 
  \hfill 
  \begin{minipage}[htb]{.47\textwidth}
    \includegraphics[width=60mm]{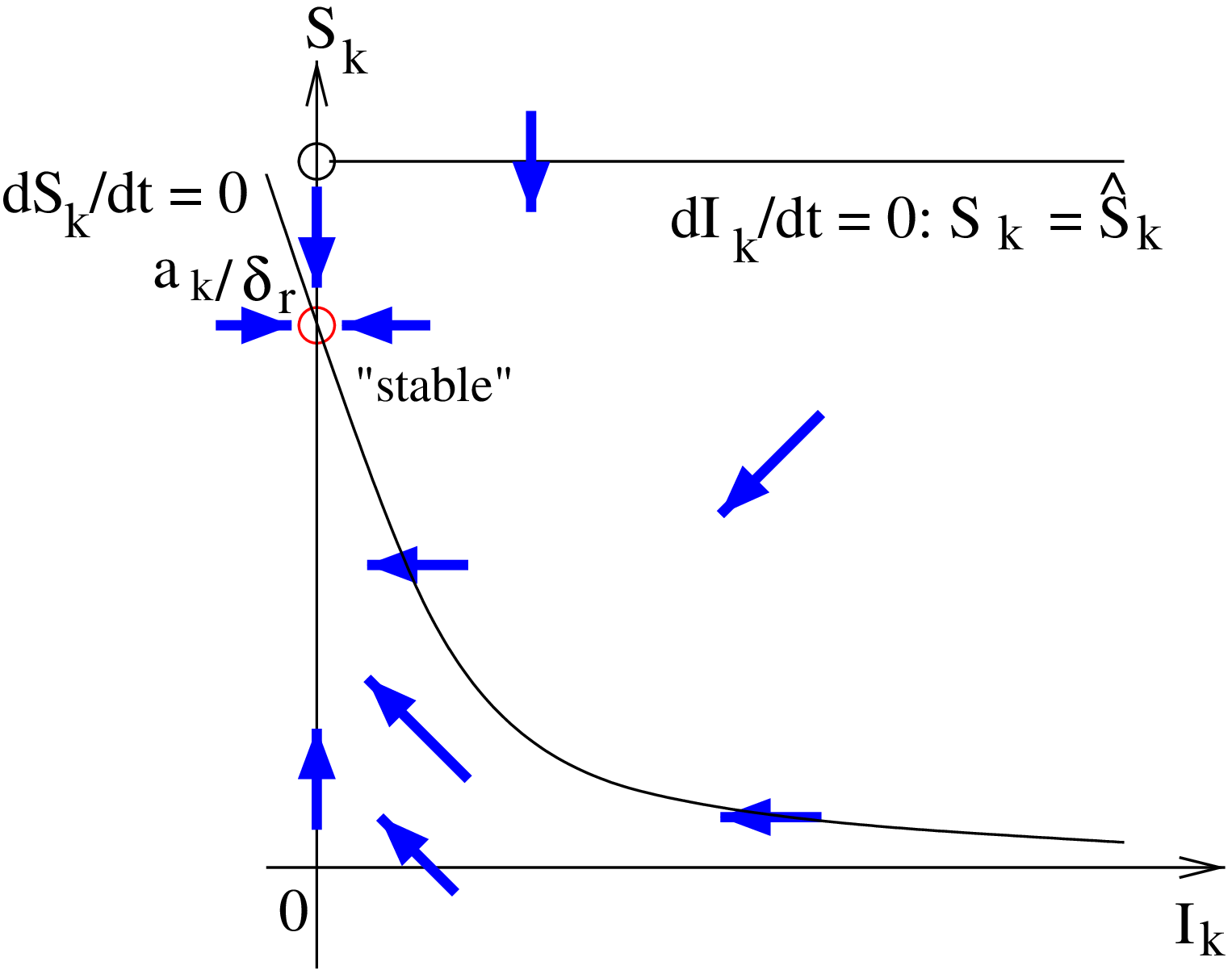} 
    \begin{center} (b) \end{center}
  \end{minipage} 
  \caption{Saddle-node bifurcation between  
	(a) damped oscillation of recoverable prevalence and 
	(b) convergence to the extinction 
	by the immunization in the heterogeneous SIR model.
	The state space is changed by the bifurcation parameters
	$\delta_{r}$ and $a_{k}$ for the value of 
	$\hat{S}_{k} \stackrel{\rm def}{=} 
	\frac{\delta_{0} + \delta_{r}}{k b c_{k}}$.}
  \label{fig_bifurcation}
\end{figure}

Next, we assume $I^{*}_{k'} = 0$ for all $k' \neq k$ 
to discuss the extinction. 
On the section, the nullclines are 
\[
  \frac{d S_{k}}{dt} = 0: \;
	S_{k} = \frac{a_{k}}{\delta_{r} + k b \Theta} 
	= \frac{a_{k}}{\delta_{r} + k b c_{k} I_{k}},
\]
\[
  \frac{d I_{k}}{dt} = 0: \;
	S_{k} = \frac{(\delta_{0} + \delta_{r}) I_{k}}{k b \Theta} 
	= \frac{\delta_{0} + \delta_{r}}{k b c_{k}}, \; 
	(I_{k} \neq 0)
\]
for Eqs. (\ref{eq_hete_S_immune})(\ref{eq_hete_I_immune}).
The necessary condition of extinction is given by that 
the point $(0, \frac{a_{k}}{\delta_{r}})$ on the nullcline 
$\frac{d S_{k}}{dt} = 0$ is below the line 
$S_{k} = \frac{\delta_{0} + \delta_{r}}{k b c_{k}}$: const. 
of $\frac{d I_{k}}{dt} = 0$. 
From the condition 
\[
 \frac{a_{k}}{\delta_{r}} < \frac{\delta_{0} + \delta_{r}}{k b c_{k}}, 
\]
we obtain 
\begin{equation}
  \delta_{r} > - \delta_{0} + \sqrt{\delta_{0}^{2} + 4 k a_{k} b c_{k}}.
	\label{eq_cond_extinction}
\end{equation}
In addition, $0 < \delta_{r} < 1$ must be satisfied, 
it is given by $a_{k} < \frac{1 + 2 \delta_{0}}{4 b \gamma}$ 
from $k c_{k} = \gamma m^{\gamma} k^{- \gamma}$, 
$m \leq k < \infty$, $\gamma > 0$, 
for the generalized BA model \cite{Moreno02}.
In this case, there exists a stable equilibrium point, 
otherwise a saddle and a stable equilibrium point
as shown in Fig. \ref{fig_bifurcation}(a)(b).
The state space is changed through 
a saddle-node bifurcation 
by values of the growth rate $a_{k}$ and the immune rate $\delta_{r}$.

For the hub immunization \cite{Dezso02}, 
$\delta_{r}$ is replaced by $0 < \delta_{h} k^{\tau} < 1$, $\tau > 0$,
e.g. $\tau = 1$ as proportional immunization to the degree.
We may chose 
the $\frac{1}{k^{\tau}}$ times smaller immune rate $\delta_{h}$ than
$\delta_{r}$ for (\ref{eq_cond_extinction}).
In other words, 
the necessary condition of extinction in (\ref{eq_cond_extinction})
is relaxed to 
$a_{k} < \frac{m^{\tau} (m^{\tau} + 2 \delta_{0})}{4 b \gamma}$.
Thus viruses can be removed in larger growth rate.

The above conditions are almost fitting to the results for the
stochastic model in Section \RomanSuji{3}.
We can evaluate them using the corresponded parameters: 
$m = 1$, 
$\nu = 2 + \gamma = \frac{\gamma_{in} + \gamma_{out}}{2} = 2.2$, 
$b \leftrightarrow \lambda = 0.1$, 
$\delta_{0} \leftrightarrow \delta = 0.04$, 
$\delta_{r}$ or $\delta_{h} = 0.1, 0.2, 0.3$, $\tau = 1$, 
and $A = 60$ form 
$(\sum a_{k}) \sim \int A k^{-\nu} dk = \frac{A m^{1-\nu}}{\nu -1} = 50$.
By simple calculations, we find that 
$a_{k} < \frac{1 + 2 \delta_{0}}{4 b \gamma}$ 
is satisfied for $k \geq 2$.
The condition (\ref{eq_cond_extinction}) is satisfied for only 
$k \geq 5$ with random immunization of the 30 \% 
and $k \geq 7$ with the 20 \%, so
the extinction of viruses is difficult by spreading of infection 
from many vertices with low degree $k \leq 4$, 
whereas it is satisfied for $k \geq 3$ with hub immunization of both the 
20 \% and 30 \% by the factor of $1 / k^{\tau}$.
The delicate mismatch at $k = 1, 2$ may be from the difference of 
the complicated stochastic behavior as in Fig. \ref{fig_SHIR} and 
the macroscopic crude approximation.

\begin{figure}
  \begin{minipage}[htb]{.47\textwidth}
    \includegraphics[width=60mm]{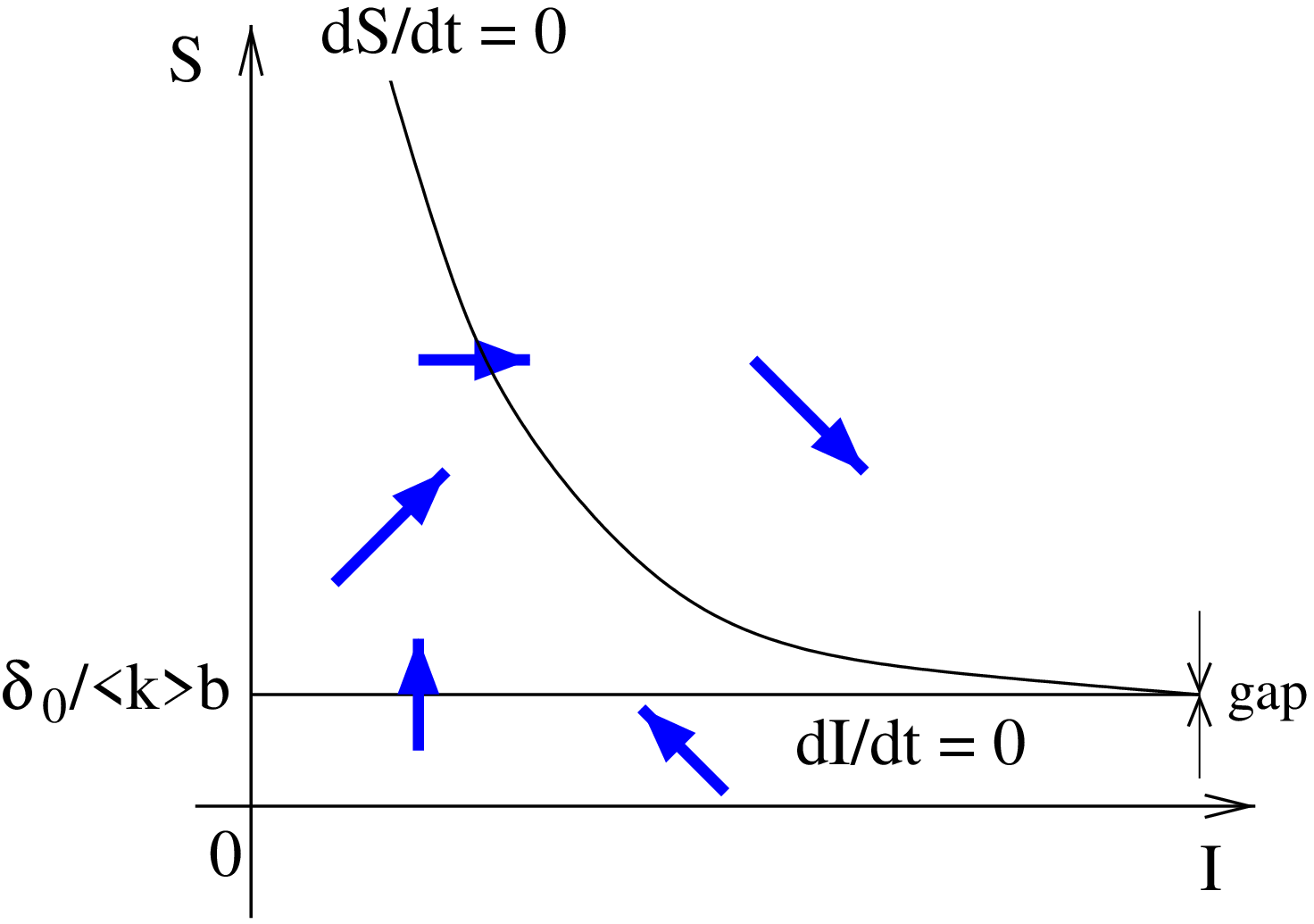} 
    \begin{center} (a) \end{center}
  \end{minipage} 
  \hfill 
  \begin{minipage}[htb]{.47\textwidth}
    \includegraphics[width=60mm]{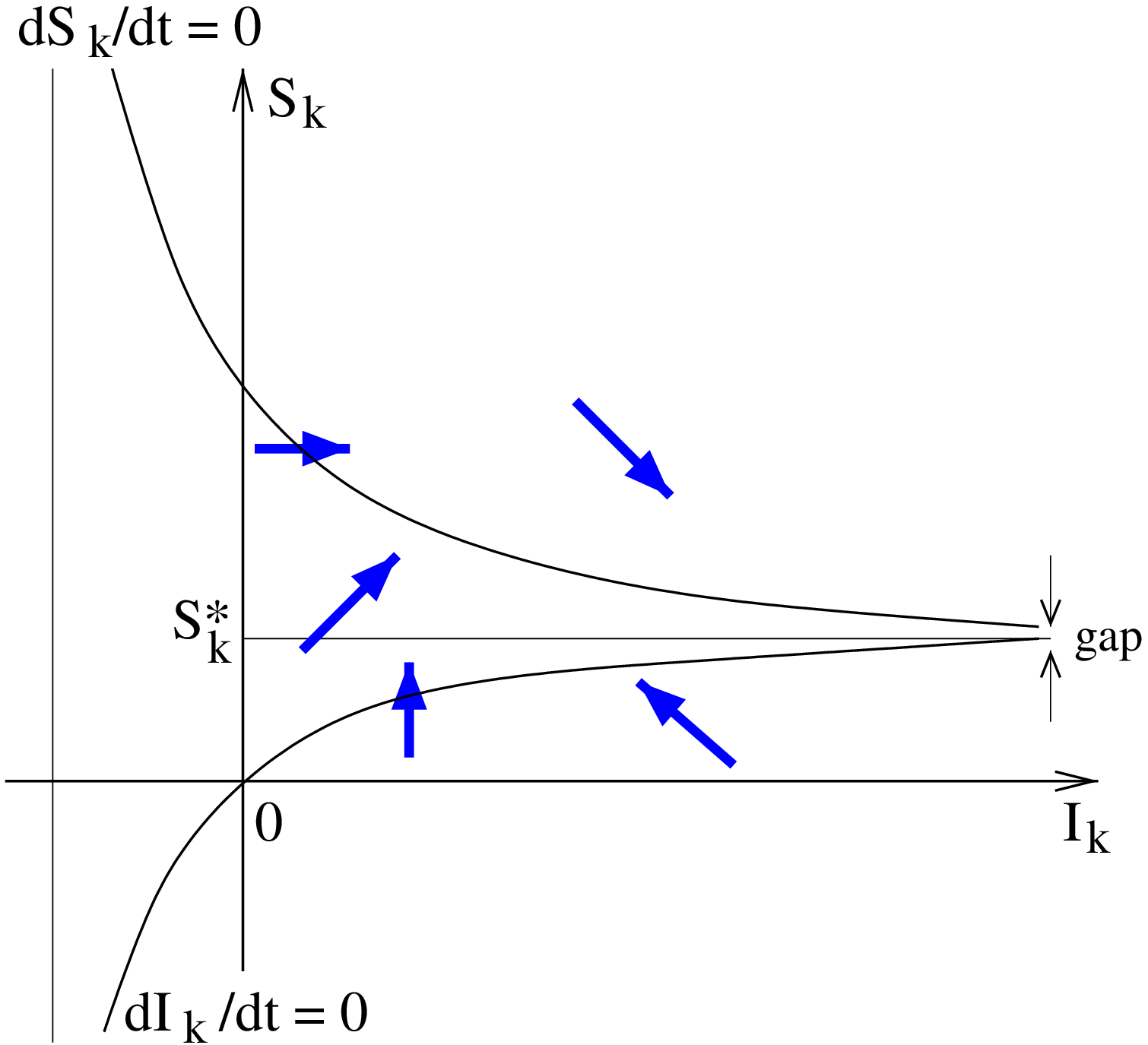} 
    \begin{center} (b) \end{center}
  \end{minipage} 
  \caption{Non-extinction in 
	(a) homogeneous and 
	(b) heterogeneous SIS models.
	The number of infected state $I$ or $I_{k}$ 
	finally diverges to infinity.}
  \label{fig_vector_SIS}
\end{figure}

\subsection{SIS model}
Finally, to show the recovered state is necessary, 
we consider the SIS models in the open system.
The time evolutions on homogeneous networks are given by 
\begin{eqnarray}
  \frac{d S(t)}{dt} & = & 
	\delta_{0} I(t) -b <k> S(t) I(t) + a,
	\label{eq_homo_S_SIS}\\
  \frac{d I(t)}{dt} & = &
	 -\delta_{0} I(t) + b <k> S(t) I(t),
	\label{eq_homo_I_SIS}
\end{eqnarray}
where $N(t) = S(t) + I(t)$.
The nullclines are 
\[
  \frac{d S}{dt} = 0: \;
	S = \frac{\delta_{0} I + a}{b <k> I} 
	= \frac{\delta_{0}}{b <k>} + \frac{a}{b <k> I},
\]
\[
  \frac{d I}{dt} = 0: \;
	S = \frac{\delta_{0}}{b <k>}: 
	\; {\rm const.}, \; (I \neq 0)
\]
for Eqs. (\ref{eq_homo_S_SIS})(\ref{eq_homo_I_SIS}).
There exists a gap of $\frac{a}{b <k> I} > 0$
even in $I^{*} \rightarrow \infty$.
Furthermore,
the time evolutions on heterogeneous networks are given by 
\begin{eqnarray}
  \frac{d S_{k}(t)}{dt} & = & 
	\delta_{0} I_{k}(t) -b k S_{k}(t) \Theta(t) + a_{k},
	\label{eq_hete_S_SIS}\\
  \frac{d I_{k}(t)}{dt} & = & 
	-\delta_{0} I_{k}(t) + b k S_{k}(t) \Theta(t).
	\label{eq_hete_I_SIS}
\end{eqnarray}
On a section $I_{k'}$: const., the nullclines are 
\[
  \frac{d S_{k}}{dt} = 0: \;
	S_{k} = \frac{\delta_{0} I_{k}+ a_{k}}{k b \Theta},
\]
\[
  \frac{d I_{k}}{dt} = 0: \;
	S_{k} = \frac{\delta_{0} I_{k}}{k b \Theta}
\]
for Eqs. (\ref{eq_hete_S_SIS})(\ref{eq_hete_I_SIS}).
There also exists a gap between the nullclines.
Fig. \ref{fig_vector_SIS}(a)(b) 
show the nullclines and the vector filed.
Thus, the dynamics in the SIS model 
is quite different from that in the SIR model.
We can not realize both of the extinction and the 
recoverable prevalence of viruses 
on the SIS model, in any case, even in the open system.

\section{CONCLUSION}
In summary, we have investigate the spread of viruses via e-mails on
linearly growing SF network models whose exponents of the power law
degree distributions 
are estimated from a real data of sent- and receive-mails \cite{Data} 
or from the generalized BA model \cite{Albert01} \cite{Moreno02}.
The dynamic behavior is the same in both 
simulations for a realistic stochastic SHIR model and 
a mean-field approximation 
without the connectivity correlations for the macroscopic equations of 
simpler deterministic SIR models.
The obtained results 
suggest that the recoverable prevalence stems from the growth of network,
it is 
bifurcated from the extinction state according to the relations of growth, 
infection, and immune rates.
Moreover, the targeted immunization for
hubs is effective even in the growing system.
Quantitative fitness with really observed virus data 
and more detail analysis with the correlations 
are further studies.

\end{document}